# Bodioid: philosophical reflections on the hybrid of bodies and artefacts towards post-human

Jiang Xu[1], Gang Sun[1], Jingyu Xu[1] & Pujie Su[1]
[1]Tongji University, College of Design and Innovation, Shanghai, China.

## Abstract

The advent of the post-human era has blurred the boundary between the body and artefacts. Further, external materials and information are more deeply integrated into the body, making emerging technology a key driving force for shaping post-human existence and promoting bodily evolution. Based on this, this study analyses the transformation process of three technological forms, namely tools, machines, and cyborgs, and reveals the construction of bodies and artefacts. From the phenomenological perspective, the essences of body and artefact existences are reflected upon, and the 'existence is construction' viewpoint is proposed. Furthermore, a technological design concept, 'bodioid', is proposed to meticulously depict the characteristics of integrating similarities and differences towards unity between the body and artefacts, based on the theoretical foundation of technology mediation and the materialisation of morality. Finally, through analogising the organizational form of language, the two key forms and specific mechanisms of 'bodioid' construction, namely 'extension' and 'mirroring', are indicated. With this in mind, the post-human existence landscape is discussed with the objective of providing theoretical insights into the study of the underlying philosophical principles of technological design.

## Introduction

Technology is progressively blurring the distinctions between information and entities, organic and inorganic material, and self-development and external design. It is transcending the boundaries between reality and virtuality and body and non-body entities, and is shaping a new social landscape characterised by diverse co-constructions. Further, the rapid succession of technologies is driving evolution, such that technical artefacts have become a crucial force in post-human self-construction.

These artefacts are embedded in the body and facilitate orderly information flow between organic bodily components and inorganic electronic devices. For example, wearable computing integrates sensing and computing technologies into everyday clothing or wearable portable accessories, functioning as body extensions (Ates et al.,2022; Weber,2010). Brain–computer interfaces connect physiological signals to external devices, revealing the brain's informational landscape (Vidal,1973). Various prosthetics such as pacemakers, cochlear implants, and retinal implants can repair or enhance human perception, movement, and cognitive capabilities, thereby restoring and improving self-survival abilities.

Additionally, technology can both conceal and reveal, as it simultaneously deconstructs and reshapes the body. Cyborgs, representing the organic integration of artificial organs with natural flesh, and electronic software, which have cognitive patterns, link internal and external spaces. This fusion generates expansive prospects that transcend inherent human attributes, bridging between the spiritual and material (Halacy,1965). Post-human subjectivity is a hybrid entity, a convergence of various heterogeneous and disparate elements, and an independent entity that melds materiality and information. Therein, the body gradually loses its physical form and is reconstructed as thoughts and information that are firmly anchored in the material world. By continuously constructing, sustaining, and reconstructing its boundaries, the post-human body shapes its existence through a dynamic interplay with its surroundings (Hayles,2000).

Moreover, computing methods are progressively being embedded in various artefacts, in the form of miniature wireless devices, positioned on our bodies and in the surrounding environment, forming a concealed information network that intricately interconnects everything to computers. These ubiquitous computing technologies offer opportunities for novel coupling among humans, machines, and the environment, thereby facilitating the emergence of new relationships (Liberati and Nicola, 2016; Satyanarayanan,2017). Intelligent environments have embedded sensors and computational capabilities; thus, they can transform interconnected information technologies into widespread practical applications (Greenfield,2010). In this way, computing technologies are being integrated into the physical environment, blending with human's daily items. Likewise, daily objects are becoming increasingly intelligent and capable of operating or making decisions in various environments, evidenced by unmanned delivery drones, autonomous vehicles, and virtual and augmented reality devices. Further examples include artificial intelligence and large language models, which are increasingly becoming vivid and 'human-like', and cyberspace, which encompasses natural, technological, and information spaces and has become a reference point in human society. We all live in a human–computer interaction network, thus becoming a part of cyberculture (Haraway,2013).

Given intelligent technology, both object and human existence forms are changing profoundly. The result is entities whose bodies are no longer innate but are acquired postnatally or constructed externally, and whose artificial objects are no longer external to their forms but embedded in them. They are and will be freely existing in, influencing, and even reshaping all aspects of our lives, including our perceptions, actions, and decisions. Human subjectivity seems to be gradually fading in the intelligent technology wave; however, it is gaining more knowledge, richer experiences, and greater possibilities for action. Therefore, we need to reflect on the nature of the body's existence and identify the appropriate perspective from which to confront the artificial entities that were once external to the body.

Consequently, as a reflective force, subject-, technology-, and objects-based philosophies are becoming entangled. This entanglement is necessary to reconsider the complex relationships between postmodern humans and objects. Therefore, this study, based on the phenomenology and post-phenomenology theory and adopting an intermediary perspective, re-evaluates and develops philosophy of body and technical artefacts, and explores the inherent characteristics of mutual construction and shaping between 'artefacts' and 'bodies'. Specifically, using the 'bodioid' (from English *body* and *-oid* 'resembling', meaning new hybrid forms of bodies and artefacts) concept proposed here, this study establishes an intermediate and generative philosophical framework for technical artefacts.

## From Tools and Machines to Cyborgs

Human existence characteristics reside beyond the body (Sartre,2007). Life constitution is not only based on human intentions or concepts but also on the arrangement of material artefacts or substances, as well as exchange and interaction with the world. Human subjectivity resides in self-construction through perception and engagement with objects. Along rational trajectories, perception, action, plan, design, and construction are employed to project ourselves, shape our destinies, and transform the external world to realise our intentions. Thus, through technology, humans progressively alter objects' functions and forms, transitioning them from tools and machines to intelligent human–machine interfaces. The diverse potentials of technical artefacts are collected in the body schema, thereby altering specific perceptions or action perspectives (Black,2014). Simultaneously, new experiences generate ambiguous subject–object relationships, from detached observation of interactions between things to the realisation of one's entanglement in them. Consequently, human existence is embedded in mutually constructed relationships, and the body–objects boundary gradually fades.

Tools provide a starting point for exploring body fluidity, as humans and tools mutually invent each other (Stiegler et al.,1998), and are touchpoints for body–environment interactions. Humans use tools to reshape the boundaries between the body and the environment by infusing in them humanistic traits through language and technical symbols. Therefore, tool use as human organ extension results from humans externalising their characteristics into the natural world, and it reflects an anthropomorphistic nature. The essential existence of humans and technology is constructed during the process of design and use, and this inherent identity determines the essence of 'being-with' (*Mitsein*) between humans and tools.

If tools are considered appendages of the body, machines represent external transcendence. With their accuracy, standardisation, and continuity, machines promote changes in the social division of labour and labour practices, and shape new human-machine relationships. When human labour is replaced by natural force, the rules derived from experience are correspondingly replaced by natural sciences, and the form of labour materials determines the form of technology (Marx,2004). Automation comes at the cost of the disappearance of meaning, the value of labour is explicitly quantified, and the misplacement of subjects and tools leads to the loss and duality of subjectivity in the process of objectification. While humans create artificial systems, artefacts construct the human practice process. The essence of alienation is also an extreme construction. Machines provide new possibilities for action while also partially replacing human behaviour. Workers' activities become alien activities that do not belong to them. The labour machine controls their actions and incorporates them into its own system, and human beings become a means rather than the purpose of machine labour, constituting the alienation of the labour subject. The diversity of artificial forms and functions, coupled with the mass production and standardisation of machines, provides more possibilities for action; however, the human body and cognition have fixed capacities. Therefore, an inherent conflict exists between increasingly complex functions and smaller operating interfaces. For users, machines become black boxes that conceal internal technology and structure, and they can only be operated based on instructions written on their surface. Meanwhile, for designers, machines are the materialisation of production sequences and technological principles. This is the separation of the design and use contexts.

Whether 'human is machine' or 'machine is human', the underlying assumption is to view humans and machines as separate poles, where their initial opposition and similarity reveal an internal mutual construction and potential unity. Perspectives such as 'organ projection, extension, or substitution' reveal the similarity between machines and humans, with humans continually creating themselves through tools or machines (Kapp,2018). Intelligence development has gradually transformed machines into humanoid entities. The characteristics of the subject have gradually permeated machines, and control has evolved into interaction, which has alleviated human alienation to some extent. Technology is forming a kind of humanity, and people silently use it to construct new selves (Borgmann,1984). The substitution of power with intelligence provides more opportunities for machines to integrate with the body or into users' daily lives through intelligent and miniaturised technology, such as artificial intelligence, the Internet of things, augmented/virtual reality, and advanced human–machine interfaces. Intelligent machines act more like tools, breaking the spatiotemporal division of work, entertainment, and social life created by production machines. Their flexibility, intelligence, and transparency allow them to gradually integrate human characteristics into machine usage, concealing the complex structures behind easy-to-operate and concise external interfaces. Furthermore, human intelligence characteristics are imitated and deeply integrated into the structure and function of machines, and machines are embedded within organic bodies, forming cyborgs. Micromachines and advanced information technologies are closely integrated, implanted, or placed in the body, allowing the body to transcend physiological limitations, enter the realm of omnipresence and omnipotence, and possess sovereignty over all things. With the assistance of micromachines, it has become possible to perform tasks that are otherwise challenging for a traditional body.

The transition from tools to cyborgs is fundamentally a process of mutual construction between humans and artefacts. It progresses from a state in which tools are unidirectionally dominated by humans, through the era in which humans become observers and assistants of machines, and finally towards the integration and mutual shaping of both entities. Mutual construction is rooted in two heterogeneous forces: similarity and difference, and self and others. These forces, while simultaneously attracting and repelling each other, are connected by inherent similarities and separated by different forms. Through continuous struggles, they gradually merge and eventually form a homogeneous whole. This bidirectional and convergent construction process leads to a state in which objects are gradually becoming bodioid artefacts as we become post-humans. Thus, the boundary between the body and artefacts steadily fades. Artefacts, such as artificial hearts or brain–computer interfaces, whose form and function are attached to the body, become extensions of our perceptual, cognitive, and operational capacities. The subjectivity of human beings is reconstructed by artefacts, which essentially reveal specific forms of self-reflection in self-discovery. Consequently, a shared subject consisting of the self, others, and artificial objects gradually forms. This transformation involves the reconstruction of self–other and private–public relationships through which artefacts become tools for both collective and individual self-expression. Furthermore, virtual and reality domains, as well as information and matter, are gradually integrating. In this context, pattern is more important than presence, and information is more dynamic and essential than material form. The information and real worlds become intertwined. Computer simulations place the body within a feedback loop involving a computer-generated landscape, where some activities occur in real life and others in virtual reality. Thus, technology plays a mediating role in shaping our understanding of the world and influences our modes of existence.

## Existence is Construction: The Phenomenological Basis of Bodioid

Exploring the isomorphism between artefacts and the body requires breaking away from the binary narratives of consciousness and object, body and world. It requires a return to the perceptual experience of the 'embodied body', because 'we are the bodies, we exist in the world through our bodies, and the bodies evoke experiences of the world presented to us' (Merleau-Ponty,2002). Existence originates in the body, and construction is based on the body's construction. We use artefacts to emulate our bodies and design our existences. During the process of material cultural changes, the forms and skills associated with objects are sedimented in and become integral components of our bodies. The body and artefacts inherently have fundamental similarities. Similarity is an 'inherent, silent, and natural universality that connects many individuals naturally' (Marx, 2016), which is more of a relationship than an entity. It delineates the ongoing conflict and coordination between identity and particularity, reveals hidden commonalities in the existence of individuals with certain characteristics, and constructs a delicate balance between identity and difference. Phenomenology is the study of the structure of human consciousness from a first-person perspective. The manner in which things are experienced, and the orientation and grasp of consciousness towards things receive due attention and description (Gallagher,2012). This provides new possibilities for analysing the similarities between the body and objects.

**The body exists as a species-being.** Humanity is a species existence, and the self-conscious practice of activity is the fundamental mode of human existence (Marx,2016). As common subjects, we co-exist in natural and social environments, or collectively face objects under specific conditions (Sartre,2007). Nevertheless, humans possess the concept of 'self'—the individual, who, due to the unity of consciousness, above all changes he may encounter, remains the same individual (Kant,2017). Subject, that is society. The essence of the species-being emerges and develops in the objectification of the free class essence in human culture; that is, it can be accumulated in tools and objects, habitual systems, and language. The species-being is 'a class that is reproduced under various cultural conditions and is self-conceptualised in the formative process' (Korsgaard,2018). In species activities, humans and the world belong to and interact with each other, and unity of negation is achieved between the subject and object. On one hand, real individuals transform the inorganic world through their social and historical practical activities, create an objective world, and realise subjective purposes in an objectified manner. On the other hand, in this practical activity, the objective natural world, as the means, object, and tool of human life activities, is transformed into the 'inorganic body' of humans, which becomes an internal link and organic component of human life activities and the living world. The meaning of existence is constructed through subject–subject communication and subject–object interaction.

Subjects make efforts to dissolve distorted communication, share experiences, and convey meaning among themselves, which connects numerous subjects and achieves unity with the 'non-self' in a harmonious and consistent state. Artefacts facilitate these encounters between subjects. Subjects and objects strive to deconstruct the 'concealed interaction' between them, achieving a fusion with the 'external' through mutual approaches. Thus, the connotation of the species subject and the phenomenological body have an inherent unity: the body in which one exists is both a physical body and a self-identifying entity within self-consciousness and among subjects. Simultaneously, it possesses an identity of existence in the world.

**Artefacts are the unity of class and individual realities.** Technical artefacts are material objects manufactured under certain technological conditions to achieve specific purposes. They are durable existences resulting from the transformation of nature by humans and emerge as relational entities based on collective consciousness and rational practices. The unity of intentionality, function, and structure constitutes the inherent reality of technical artefacts (Mitcham,2002,2006). Low-order artefacts represent foundational reality, such as physical or scientific reality, corresponding to intentional input phases. High-order artefacts represent functional or technological realities corresponding to the intentional output phase. Functional intentionality is externalised through a specific artefact from the predesign stage, revealing the artefact's external properties. At this point, the internal and external natures of the artefact are unified, demonstrating unity between abstract functional and specific physical entities. The meaning of an artefact can only be revealed through its use (Heidegger,2010). Further, the unity of class and individual realities can only manifest naturally in artefact–human interactions. The existence of a technical artefact is always dependent on a holistic technological concept, and as part of the whole, it serves as a 'thing for the sake of doing...'. Technical entity usage promotes connections between individuals and reality. Although the ready-to-hand artefact itself may be absent from perception, it also plays a constructive role and mediates the people–world relationship. Therefore, as individual reality, an artefact can only be integrated into social culture through the user's body, revealing its inherent functional significance. As class reality, artefacts can only diffuse from individual to collective experiences, possessing cultural and historical conceptual significance. Furthermore, the directedness of artefacts is uncertain. Artefacts are constructed during the design process, presenting the collective intentions of users in the design and promoting encounters between designers and users in different life scenarios. Construction refers to the co-construction and integration of designers, users, artefacts, and usage environments. When examining 'user–artefact–world' relationships, the variability of artefacts, users, and the environment (concerning their physiological characteristics, cognitive abilities, perceptual capacities, and social attributes) should be recognised. Artefacts themselves exhibit different functional states in various scenarios, and the usage environment of artefacts varies among designers, manufacturers, and even users who are unable to fully control them.

**Meaning originates from living experience.** Subject–subject communication and subject–object interactions occur throughout the world. Modernity strictly distinguishes between subject and object, and between the mind and its existence in reality, with the subject becoming the benchmark for reality. Reality is the objectification of what the subject gazes upon. The world becomes a representation of an object outside a 'given' world, projected onto the back wall of the human consciousness darkroom. Phenomenology, on the other hand, aims to describe how things manifest in our conscious experience, with the subject being a live entity existing in the world and perfectly aligned with consciousness and the body (Merleau-Ponty,2002). The only way we can access things in reality is through consciousness, which is always about something formed during interactions outside and between subjects (Husserl,2012). Subjectivity is no longer self-evident, and the intentionality of consciousness is constructed through perceptual interactions with others and objects in the lifeworld. This breakthrough in the boundary and co-construction between subjects and objects in the closest lifeworld is a postmodern trend. We exist in the world as active agents intervening in plans that are defined pragmatically and socially. Structural coupling occurs between the acting subject and its surrounding environment based on an emergent action-guided meaning. An action always precedes oneself as a way of existing in the world (Heidegger,2002). Actions integrate the situations shaped by past actions and the planned future towards which they are heading, into the current context that can both limit and achieve it, where the meaning of action emerges from this internal temporal structure (Bergson,2022). The lifeworld is a collection of various situations in which we are involved — it is the world of our life, not only the world unfolding in front of us as perceiving subjects, but also the world that has already played a role as a meaningful background for all our actions and interactions. Therefore, the lifeworld is a concretization of our existence.

Meaning originates from construction: 'Dasein' and 'Mitsein' have already formed the structure of the lifeworld, and artefacts become a part of it. The construction process is embedded in the lifeworld context, in which objects naturally acquire a subject's characteristics. The properties of artefacts are acquired, and the body achieves extension and expansion with them, collectively forming microunits in the lifeworld. Thus, the body–artefact–world relationship is a co-existence form, representing the essence of 'bodioid' artefacts.

## From mediation and moralization to bodioid: the subject and object in construction

As mentioned above, human and artefact existence exhibits inherent isomorphism, and bodioid artefacts are constructed through continuous interaction. Then, from this perspective, how does this existence unfold? The flesh serves as the primary mediator and interacts directly with the external world. The 'inner self' participates in the world construction process during sensory experiences, which constitute the most fundamental form of human cognition and practical activity (Merleau-Ponty,2002). This mediation, which originates from modified natural objects, distinguishes human activity from that of other species. The subjects of the activity create mediations from objects, not only causing the differentiation of objects, but also projecting themselves onto the mediations, making the objects mediational subjects. Thus, mediations possess both subject and object characteristics. For the object, mediations transmit the actions of the subject, thus becoming an extension of the subject's limbs or organs. For the subject, mediations convey the reactive actions of the object, effectively serving as an object that directly interacts with it. Mediations simultaneously signify and negate a certain form of indirectness between the subject and object. They not only separate the subject and object, but also connect them in a specific way, and this unity of similarities and differences becomes the most fundamental form in describing bodioid.

**'Body-world' perceptual pattern.** Artefacts link humans and reality (Heidegger,2010). The world itself exists as a perceived object, and the 'I' perception actualizes it; therefore, 'I' am the expression of the world itself. The body exists in the world with perception and action. This unity of perception and action is a unified form of pre-cognition and pre-objective. The phenomenal space constructed through the body lays the foundation for body perception. The body is the most primitive space, and the external space can be created only by projecting 'content' and 'meaning' outwards based on the body space. The body can occupy external space through its activities, thus expanding its living space. In turn, external space gives the body space a form of self-evidence (Merleau-Ponty, 2002). For example, for the blind, a cane is not just an object, but an extension of their perceptual abilities and range, taking on the function of a visual organ. This demonstrates that technical artefacts can extend the body's perception and space.

**Perceptual mediation.** The perceptual body–technology–world approach initiated by Ihde (1990) focuses on the perceptual and hermeneutic implications of technologies by analysing how specific perceptual techniques facilitate the experience and interpretation of reality. Embodiment theory is equivalent to Heidegger's ready-to-hand concept. Technology, in its mediational position, becomes transparent while highlighting its own power, concealing and revealing simultaneously, which makes human beings and technology the same subject jointly pointing towards the experienced world. Hermeneutics provide another path for technology's entry into reality. Technology presents the referred world through texts, and through reading the text of technology, human intentionality points to the world reinterpreted by technology. 'Embodiment' and 'Hermeneutics' are two key forms of technology extending the body. Embodiment emphasises the fusion of perceptual experience and the expansion of the body schema, while hermeneutics emphasise the coupling and representation of knowledge. Through text and data, humans are connected to machines, with technology here being an extension of the cognitive reconstruction of real-life experiences. 'Alterity' describes the relevance between humans and technology-as-other. Technology has its own development environment and usage requirements, leading humans' attitudes contradictory. They wish technology to be more than just a tool but also desire it to remain somewhat different from humans. This is caused by machines progressively mirroring others. 'Background' explores the technologies embedded in the background. The connection and expansion of devices constitute machine systems, and 'absent' machines as a whole construct the technological background beyond humans, seemingly projecting our dreamy scenes and needs into the real world.

**Action mediation.** Technology appears to 'act' in the world in a way different from humans, effortlessly crossing the modernist subject–object boundaries. In the lifeworld, 'ambiguous objects' exist between subjects and objects, and these cannot be reduced to either subjects or objects. To understand them, an examination from the subject's action perspective is necessary (Latour,1987). Our reality is formed by the constant interactions and translations between humans and non-humans, which form ever-new connections and entities. Artefacts create 'meetings' and 'conversations' between users and designers, with close care and empathy written in scripts and unforeseeable forms constructed in interactive scenarios. According to actor–network theory, time, space, and action overlap, and goals and means are coupled in specific contexts. 'Technology and its actions occur simultaneously, confined to the most basic interactions of humans... Without the mediation of technology, humans cannot exist decently' (Latour, 2012). Technology is not merely a mediator that helps humans achieve their intentions in the material world, but also a regulator that actively promotes the shaping of reality. Technology both provides the means to and promotes the construction of new purposes. It does not provide functions but enables detours and is endowed with meaningful living characteristics in its functional associations.

**Materialization of morality.** Ethics is no longer a framework for subjects' elusive reflections and moral judgments but also includes the moral design and actual usage experiences of material entities. Designers materialise morality, and technology has transcended being a simple neutral tool of convenience. While fulfilling its functions, technology also frames our actions and world experiences, thus actively participating in our lives (Verbeek, 2006). The subject is no longer inherently predetermined, but rather constructed, embodying a non-autonomous existence and a passive product of power relations (Nietzsche,1968). With the help of such power, subjects develop a form of free relationship. Ethical discussions focus not only on what we should do as subjects, but also on what kind of subjects we intend to become (Foucault,2005). Materialised moralities emerge in their association with users and are both neutral mediations and active subject builders. Moralising technology is both a means to and purpose (Latour, 1987) for shaping the actions and decisions of subjects through invitation and inhibition. Intentionality should not be merely a human trait, but one of technology as well. Freedom should not be understood as the absence of external influence on actors but a practice of dealing with this influence or regulation as well. Artefacts are interpreted, regulated, and defined in their usage context, acquiring roles and identities through their interrelationships with users. The materialisation of morality constructs mixed class subjects: (1) human subjects who perform actions or make moral decisions, thus determining the interpretive meaning and usage form of technology; (2) designer subjects who embed specific regulatory forms, thus influencing the final mediation role of technology; and (3) technological subjects who sometimes regulate human actions and decisions in unpredictable ways (Verbeek, 2005). This represents a quasi-subject in which heterogeneous elements are harmoniously intertwined by internal associations, conflicts are smoothed out in interactive contexts, and a coordinated and unified whole is constructed. The purpose of moral self-construction is not to rescue humanity from technology but to harmonise technology and humanity satisfactorily.

**Towards bodioid artefacts.** The construction foundation of class subjects is the phenomenological body, which serves as the basis for the intrinsic similarity and association between designers, users, and artefacts. The phenomenological body extends its presence through practical engagement with the lifeworld, incorporates external objects into its body schema through intentional projection, and builds the foundation of meaning transmission between subjects through the body of 'Mitsein'. Artefacts are coupled to the body's functional intentions and material realities. Their forms and functions depend on their associations with the body in specific contexts. Only when closely approaching the body—in terms of form, function, or distance—and jointly undergoing experiences can artefacts flexibly enter the lifeworld and can their existence's significance be revealed. The concept of artefacts is rooted in bodily experiences and linguistic logic, acting as a mediator of intentions or as part of the body. Knowledge about and experiences of artefacts are stored within the body, which serves as source of use and design. Based on this, this study introduces the 'bodioid' concept to describe the mutually constructing, constantly approaching, and gradually unifying relationship between the self and the external world. This concept offers an interpretive perspective on how artefacts merge or project onto the body, and how the body internalises experiences and knowledge about artefacts. It also reflects on how to shape artefacts to achieve 'self-practice' and reshape the common subject. Thus, intermediate and generative technological design thought has emerged as a new force in self-construction and world transformation.

**Bodioid construction.** The bodioid construction concept is a characteristic of the subject–object–subject relationship, which is constantly changing, mutually mediating, and tending towards uniformity. A bodioid artefact is the objectification of this construction progress. The following four bidirectional construction processes between humans and artefacts are revealed through bodioid artefacts. (1) Technology is transparently embedded in the body, directly shaping it, and even becoming an intrinsic part of it. (2) Technology is inherently integrated into the body schema, becoming a part of humans' intentions towards the world, not just the object of intention. (3) Artefacts become quasi-others and independent agents with human-like thoughts and actions. (4) Technology constructs the body into virtual information flow, directly intervening in the system of artefacts, where real virtuality replaces the physical presence, and the patterns/randomness of information substitutes for the presence/absence of flesh.

The bodioid concept reflects a construction of existence. Human existence is contained within human bodies. The living body existing in the world indicates the form and change of existence, the flesh's occupancy of the world indicates its spatial existence, and intentionality pointing to everything in the world signifies its meaningful existence. The essence of existence extends over time, manifests changes in human perception and action, and finds a unified meaning within the infinite possibilities of change. Artefacts exist in their functions and structures (Kroes and Meijers,2002), which can be understood in two ways: what they can be used for (physical structure) and what we want to use them for (intentional function). Artefact design and evolution are not only about physical structure and performance adjustments but also changes in human intentions and needs, through ultimately seeking to make artefacts more efficient, effective, and suitable for us, as are our bodies. By echoing this objective, bodioid construction not only acknowledges the past but also looks towards the future, while continuously reinforcing and deconstructing both.

Bodioid construction also describes the heterotopy between reality and virtuality, primacy and creation (see Fig. 1). On the one hand, as technology and the body fuse, jointly understanding and transforming the world, a bodioid mediation that is reflected in the similar-to-self nature of the body–technology–world relationship is formed. On the other hand, the new body authorises artefacts to become agents of action, replacing humans in establishing connections with the external world. This body lives in a world presented and constructed by bodioid objects, also reflected in the similar-to-others nature of the body–technology–world relationship. Bodioid artefacts encompass two self-representations: the pre-reflective self and the narrative self. The living body can integrate things into the body schema and extend them into the surrounding environment beyond the natural body's ability. Self-identity is not only a product of introspection but also the result of intersubjective conversation. Bodioid constructions offer a new narrative mode centred on the body, in which designers and users meet within the artefacts. Thoughts and memories are formed through shared consciousness and temporal extension generated by the self-constructed narrative, thus reshaping personal moral worlds and connections to the entire social world through phenomenological embodiment. Personal moral worlds and connections to the social world are reconstructed through bodioid artefacts.

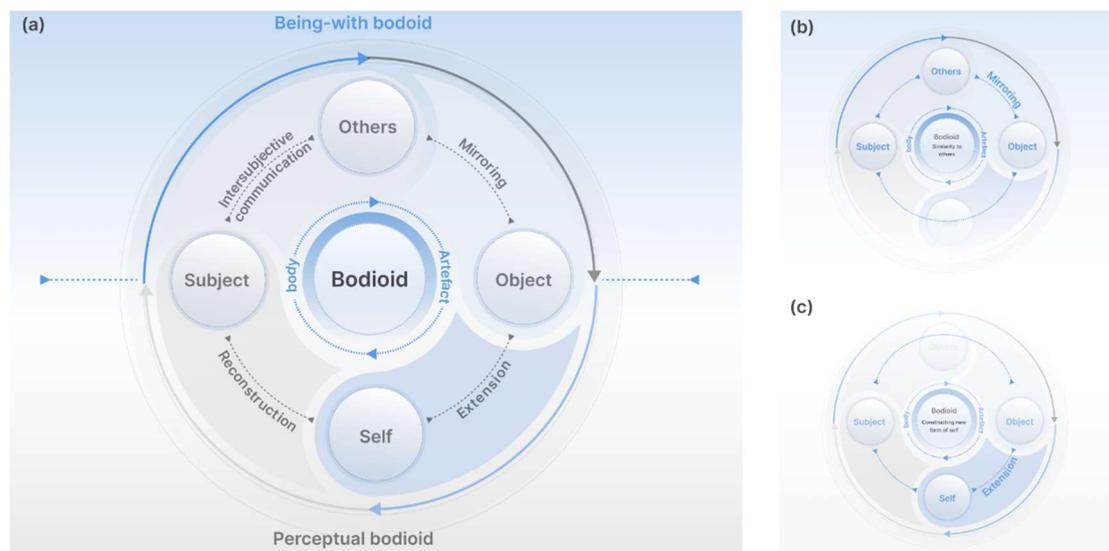

**Fig. 1 The bodioid construction in various subject-object-subject relationships. a**, the construction between two dimensions, subject-object and self-others. **b**, the similar-to-others nature of bodioid construction. **c**, the similar-to-self nature of bodioid construction.

## Extension and Mirroring: Key Forms of Bodioid Construction

**Language: a natural bodioid object.** Objects and language have natural similarities, and both interpret the existence of humans to some extent. Language blends with a subject through practical engagement and becomes a communication and expression organ (Merleau-Ponty,1968). Words refer to the naming of objects, and the organisation of words by the subject is similar to the use of ready-to-hand tools. The body can also create language expressions according to intention, without conscious thought. Meanwhile, language guides and shapes human thought fluidity, giving momentary inspirations a concrete form through the organisation of sound, symbols, or grammar. The conceptual world is constructed by language, allowing the primordial body to obtain depth, which marks the beginning of the separation between humans and original things within intention to gain spiritual autonomy. Therefore, while using language, humans also construct its meaning and usage, creating a co-existence on both the subjective and cultural levels (Heidegger,2002). Language, as an expression of the body, has components beyond the primordial body and is an extension of the body schema, possessing independent quasi-corporeality. Thus, language is an expression created by existence, which is a type of bodioid object shaped by the subject, approaching the body, transcending the body schema, and constructing a new existence.

Therefore, what is the organizational form of language? Similarity organises the operation of symbols, making humans aware of many visible and invisible things and guiding the art of representation. Representation appears as repetition: the stage of life or the mirror of the world, which is the identity of all languages (Foucault,2005). The following four forms and connotations of similitude are proposed:

*Convenience:* This form strongly denotes the adjacency of places, as they come sufficiently close to one another to be in juxtaposition. A resemblance that becomes double: resemblance of the place, and similitude of properties. …In the vast syntax of the world, the different beings adjust themselves to one another. …This convenience that brings like things together and makes adjacent things similar, the world is linked together like a chain.

*Emulation:* A sort of convenience that has been freed from the law of place and is able to function, without motion, from a distance. … There is something in emulation of the reflection and the mirror: it is the means by which things scattered through the universe can answer one another. …The links of emulation form series of concentric circles reflecting and rivalling one another.

*Analogy:* Convenience and emulation are superimposed in this analogy. Like the latter, it makes possible the marvellous confrontation of resemblances across space, but it also speaks, like the former, of adjacencies, of bonds and joints. …Through it, all the figures in the whole universe can be drawn together. There does exist, however, in this space, furrowed in every direction, one particularly privileged point: it is saturated with analogies. This point is man.

*Sympathy:* Sympathy is a principle of mobility. It inspires the movement of objects, and even the most distant things can draw near to each other. By drawing things towards one another in an exterior and visible movement, sympathy also gives rise to a hidden interior movement — a displacement of qualities that take over from one another in a series of relays. It is an instance of the same so strong and insistent.

Bodioid construction reveals the constructive role of body–artefact interaction, the starting point being their similarity and the process being the confrontation and fusion of two heterogeneous elements. The living characteristics are reflected or extended into artefacts; therefore, the purposes of design and use point to human intentions and needs. Artefacts magnify or diminish perceptions, invite or inhibit actions, and guide certain moral norms, thereby signifying latent skills or knowledge within the body. These characteristics constitute the essence of human beings. Similar to language, bodioid artefacts are mainly generated through two forms: extension and mirroring (as shown in Fig. 2). Extension originates from convenience, whereas mirror originates from emulation. Analogy and sympathy are key forms of self-organisation for both.

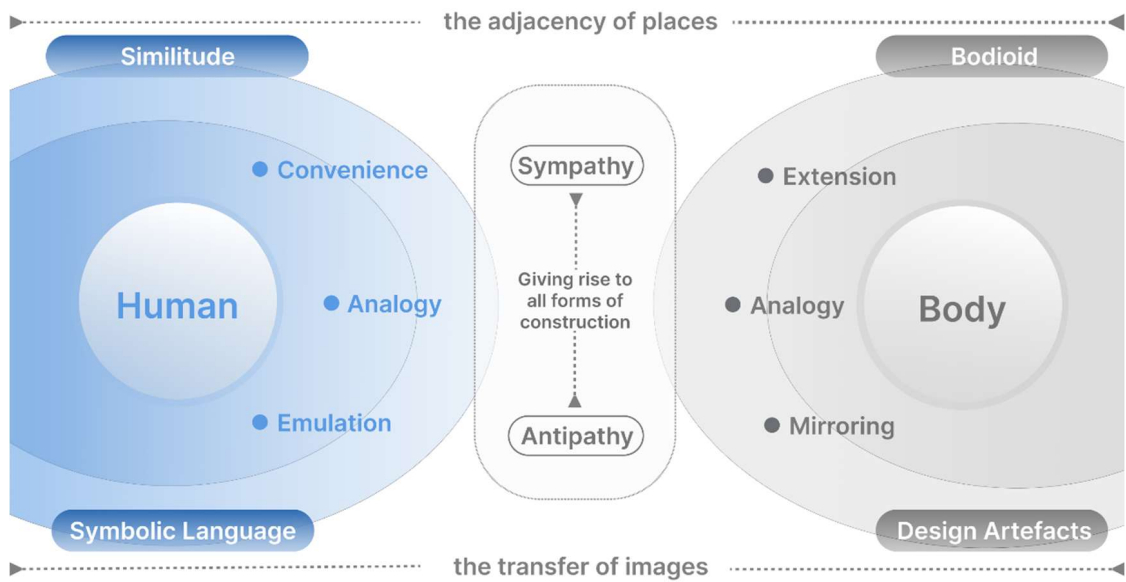

**Fig. 2 Two key forms of bodioid construction.**

**Extension is similar to convenience.** Ordinarily, as described previously, extension entails the continuous construction of bodily existence, perceptual experiences, action possibility, and cognitive characteristics through a tool juxtaposed with the body schema, leading to action. Similarly, humans design and use bodioid artefacts based on the extension of their bodies, which includes the bidirectional proximity and suitability of the body and objects at the interface position or functional attributes. Here, convenience extends the body schema at the skill level, while creating more interface forms at the physical level.

*Extension is a tangible replacement, compensation, and expansion of the flesh.* Body limitations compel humans to seek external tools. The initial intention of design was to supplement or replace the body's organs (Kapp,2018). Gene editing technology enables precise modification of specific genetic sequences, deconstructing and reconstructing existence for the better. The body is the most direct object of the bodioid construction, which is an inherent enhancement (Heidegger,2002). Prosthetic technologies or organoids (Zhao et al., 2022), drawing analogies from the control and sensory feedback mechanisms of natural limbs or organs, facilitate the body's restoration and reconstruction. The tactile feedback function is simulated, and lost movements are reconstructed (Shu et al.,2024).

*Extension creates new content and forms of perceptual experiences.* Perception is the starting point for understanding the world, the 'paradox of interiority and transcendence', and the foundation for unfolding all actions (Merleau-Ponty,2002). For example, blind people perceive canes as more than mere objects; they view them as extensions of their visual range and reach of touch (Ihde,1990). Similarly, machine simulation of vision involves more than simply replacing the human eye with sensors. It introduces computing systems for target recognition, tracking, and measurement, thus consciously or unconsciously transforming visually impaired subjects' life perceptual experiences (Kim et al.,2022).

*Extension creates richer possibilities for action.* The constructive effect of extension on the body changes its patterns, extends its agency, and redefines its peripheral space. We encode the distance of nearby objects through relative manipulation and touch. Technological tools can temporarily modify the spatial representation of actions by modifying the areas that the body can touch, thereby creating new functions and meanings. For example, artificial muscles simulate the biological properties of the body and the states of force in various spatiotemporal dimensions, enhancing physical strength in multiple scenarios such as healthcare, prosthetics, robotics, bionics, and micro-nano machines (Shi and Yeatman,2021).

*Extension involves the transformation of cognitive forms.* Artificial intelligence application involves humans' use of machines to help themselves understand and synthesise the characteristics of external objects and abstract features. Some intelligent machines exhibit features such as adaptability, self-learning, self-organisation, self-coordination, and self-optimisation, allowing them to be embedded into different environments so as to interact naturally with people, contribute to the creation of new life scenarios, and change the way we use and recognise products, among other functions (Mühlhäuser,2007).

*The fusion of technology and the body creates a transcendent quasi-subject.* Object experiences are inherently related to body experiences, and the two are isomorphic. The world demonstrates the universal convenience of objects, and extensions join bodies and objects to form mediations of subjects' world experiences. Human subjects inscribe morality onto technological artefacts, which can extend bodily perception and movement abilities, become the text that interprets environmental elements, and turn into the constituent components of the environment (Verbeek,2005). To a certain extent, the integration of body–technology–world experiences is achieved.

*Extension facilitates new forms of interaction and conversation between subjects.* Intelligent products interconnect with the body through various interfaces and function as organ extensions, forming a cognitive field extending into the environment. This universal convenience shapes new forms and contents of conversation among subjects, with the body as the input and output platform (Harrison et al.,2012). For example, bioinformatics-physical systems enable individuals to engage in remote voice and video communication regardless of spatial distance, facilitating emotional connections or work-related tasks. Social network platforms facilitate interaction across distance and time, and also solidify human self-expression and memory-making (Kudina,2022). Intelligent head-mounted devices provide a new information transmission channel: optical head-mounted displays present interface content, artificial intelligence delivers voice information, and users communicate with internet services through voice commands or touch gestures.

**'Mirroring' is a form of 'emulation'.** The general 'similarity' and 'fit' between objects transcend spatial constraints. The body extends itself infinitely into time and space through mirroring, while technical objects imitate the body to construct their own form and core, such that a silent similarity can break distance constraints and facilitate communication between the body and objects. Consequently, a bodioid constructs a body emulation akin to quasi-otherness, serving as an agent of our actions and objects of interaction. Simultaneously, we observe certain characteristics of ourselves through this mutual emulation. This mediation and flexible similarity between the self and others is at the core of mirroring. Bodioid construction creates a natural 'self–existence–other' mirror. Distances in time, space, attributes, and forms are folded, and two similar parties exist in constant opposition and convergence (Foucault,2005). Similarity becomes a struggle between two different forms, in which opposing parties control or construct each other, and similar objects accommodate similar objects. The mirroring constructs a series of concentric circles centred on the body, reflecting and competing with each other, and the system of objects is distributed within them.

Mirroring begins with the replication of oneself and others. For example, we understand and construct the external world on our own scale, and we shape objects similar to us to fulfil our internal needs and forms of existence. We shape objects similar to others for communication, conversation, and interaction, forming individuals of quasi-otherness. A certain similarity rooted in the existence of the body maintains the difference between reality and virtuality, and the self and the external world. We often indulge in the experience of incarnation, thinking that it is our own experience and assume that the intelligent machine in conversation is like another similar to us. This sense of belonging and similarity is often interrupted by an occasional sense of difference. Mirroring, a form of replication, decomposition, and recombination of given forms within a modified and curved spacetime, is an ectopic representation (Foucault,2005). Meanwhile, construction is bidirectional, and the mirror image has mutual similarities. We see ourselves in others and others in ourselves. We construct artefacts similar to ourselves, while our experiences, concepts, and skills regarding these artefacts reshape our bodies. The bodioid mirror construction function includes the following forms.

*The body becomes the source of technology emulation.* The essence of technology is nature emulation, beginning with the body's physiological functions. Technology compensates for instinctual deficiencies and provides more possibilities and openness for humans to adapt to external environments. The body, as the most direct object of technological emulation, becomes the centre of association between things, surrounded by similarities. Brain-inspired computing draws inspiration from the nervous system and information-processing methods of the human brain at multiple levels, including program models, system software, and neuromorphic devices, to construct a virtual machine brain based on neural network computing models (Zhang et al.,2020). Through brain-computer interaction technology, scientists aim to create a 'super brain' that integrates virtual and biological ones, ultimately establishing a new computational structure and intelligence form. Whether it is intelligent robots in the real world or virtual humans in cyberspace, most of them have an 'emulational body', imitating the entire body form, structure, and function. Today, the development of digital technology is guided by the body, creating and playing productive roles based on its structure and characteristics (Shilling, 2012).

*The body authorizes artefacts, making them agents of activity.* Technological regulation shapes our agents in terms of 'action' or 'transformative' capabilities, making them mirror images of our bodies and redefining our ability to act. Therefore, agents are not just subject, object, or system properties, but rather the ongoing reconfigurations of the world, an enactment. In emulating the body, the perceptual and active capabilities of the subject are mapped onto artificial objects. We shift from having polishing tools to being in symbiosis and competition with intelligent machines, and from inter-subjective communication to interaction and dialogue with intelligent agents. Mirroring maps human attributes to artificial objects, and as agents, artefacts shape the forms of our existence in interaction, with the emulation of perception and action being a critical form. Intelligent agents constitute multi-agent systems that operate in a shared environment, connecting virtual and real environments, emulating the body, and exploring the environment to discover possible actions. Affordance explains the principles of interaction: the relationship between an object's attributes and an agent's capabilities determines how an object may be used (Norman, 2013). Action emerges from subject–environment interactions; agents emulate the body, and action–environment relationships determine behavioural possibilities.

*Mirroring depicts objects-to-quasi-otherness transformations and subject-object to intersubjectivity transformations.* The essence of Dasein lies in Mitsein (Heidegger,2002). All objective aspects draw their ontological significance and validity from the *a priori* intersubjectivity of the subject. By emulating others, bodioids create similarity between the perceived object and the self (Merleau-Ponty,2002), reshaping the experiential pattern of physical entities and forming an interactive embodiment related to the intrinsic intersubjectivity connections. Material or virtual entities become 'living' quasi-others, not just external objects with specific functional structures and behaviours. Bodioid artefact experiences are advancing from 'grasp' in perception and action to 'empathy' in situational interactions. The essence of conversational agents is emulating human language and its implicit thinking patterns, which form a system that imitates human dialogue using communication channels, such as speech, text, facial expressions, and gestures (Pinxteren et al.,2022). Language is bodioid, and large language models (LLMs) and generative pre-trained transformers (GPTs), trained on vast amounts of text data, have been shown to be increasingly capable of performing sundry tasks that require mathematical, symbolic, common-sense, and knowledge reasoning. This is an emulation of human intelligence mechanisms, forms, and the body's environment, including language and the knowledge, culture, and environmental elements it encompasses (Lily et al.,2023).

*Mirroring reflects others while also constructing the self.* Mirrors represent the location of original identification. The bodioid artefact is an agent of subject perception and action, through which our elements are mapped to form our own illusions and shadows. Simultaneously, it is an emulation of the other, through which meaning flows in the interactions between subjects. 'The other' is an essential premise for individuals to confirm 'the self'. Everyone sees themselves in others and recognises others in themselves (Sartre,2022). Mirroring constitutes a new way of looking, in which subject and object, and self and other merge together through bodily interactions. In such a bidirectional mirroring, we project ourselves onto it and recognise ourselves within it (Sartre,2007). Therefore, mirroring is a significant force in the construction of self and others. Virtual reality technology constructs mirror images of reality in the information space by projecting oneself into fantasy scenes. We can communicate with virtual intelligence entities and meet others in the form of information. Immersive experiences shape new perceptual experiences and lead to self-reshaping. Wearable body-monitoring devices, smart glasses, and other devices collect and analyse various body data, present visualised images to users, and enhance self-awareness through mirroring.

# Conclusion

Information technology has led to increasingly close body–object connections. Intelligent technical artefacts, interconnected social technology networks, and up-close and *unverborgenheit* life have gradually formed a new relationship centred on the body, covering distant life experiences. This article examined the constructive forces around the body, starting with an analysis of the forms of artefacts, from tools and machines to intelligent machines and cyborgs. It revealed that technology not only influences the body regarding form and experience, but also progressively reconstructs it at the material and informational levels. Furthermore, from the 'existence is construction' perspective, based on phenomenological theory, body and artefact characteristics in the lifeworld were explored. A new 'bodioid' concept was proposed to interpret the emerging design landscape and the co-construction and fusion of subjects and objects. Based on this, two key vehicles of bodioid artefact construction, namely, extension and mirroring, were elaborated.

We believe that bodioid construction reflects embodiment theory stemming from the constructive ideas of design, which involve the organic integration of humans and technology, as well as the self and others, to discuss the intermediate form of body–object co-existence and fusion. Embodiment is an examination of oneself and can be distinguished at three levels: the definite shape and inner ability of one's body, the skills we acquire to handle things, and the embodiment of culture, which means that the cultural world is interconnected with our bodies (Dreyfus,1992). Similarly, bodioid construction considers the inseparable constructive relationship between objects and the physical body. First, bodioid construction involves reconstructing the physical body. Technology is gradually infiltrating and deconstructing bodily characteristics, creating new forms of perception and action, and altering our world experiences and the projection of meaning. Second, technology is breaking down body boundaries, altering the methods of integrating and extending the body schema, and creating new bodily skills and experiential realities. Through extension and mirroring, it creates an experience field that constantly expands around the body. Technical artefacts can only seamlessly enter the lifeworld and generate meaning by being close to the body. Finally, bodioid construction reflects the self-organising post-human culture, in which the absence and presence of the body are replaced and supplemented by the randomness and patterns of information. To varying degrees, the body combines various heterogeneous and diverse elements that exist in a hybrid manner, continuously rebuilding its boundaries in unprecedented opportunities and challenges. Bodioid construction aims to reflect new human–technology relationships from an intermediate perspective, forming design thought centred on the constructiveness of existence. Further research should also gradually project this concept onto specific design methods and processes.

## Acknowledgements

This work was supported by the the National Key Research and Development Program of China [Grant Number 2022YFB3303301], the National Key Research and Development Program of China [Grant Number 2021YFF0900602], and the National Post-funded Project of Social Sciences of China [Grant Number 21FYSB037].

## Data availability

This article does not report any data and the data availability policy is not applicable to this article.

## Competing interests

The authors declare no competing interests.

## Informed consent

This article does not contain any studies with human participants performed by any of the authors.

## Ethical approval

This article does not contain any studies with human participants performed by any of the authors.